\documentclass[10pt]{article}
\usepackage{amsmath}
\usepackage{amsfonts}

\title{Conserved quantities for the Sitter particles$^*$}

\author{S. L. Cacciatori}

\begin{document}
\maketitle

{\small \it Dipartimento di Fisica e Matematica, Universit\`a dell'Insubria, via Valleggio 11, 22100 Como,
and INFN sezione di Milano, via Celoria 16, 20100 Milano}

\begin{abstract}
%
\noindent Recent observations of the luminosity-red shift
relation of distant type Ia supernovae established the fact that the expansion of the
universe is accelerated. This is interpreted by saying that there exists some kind of agent (called dark energy),
which exerts an overall repulsive effect on ordinary matter.
Dark energy contributes today in the amount of about 73 \% to the total energy content of the
universe, and its spatial distribution is compatible with perfect uniformity.

\noindent
The simplest possible explanation for dark energy is to assume that it is just a universal constant, the so
called cosmological constant. This would mean that the background arena for all
natural phenomena, once all physical matter-energy has been ideally removed, is the de Sitter space time.
Thus, the Poincar\'e group should be replaced by the de
Sitter group, and one is naturally led to a reformulation of the theory of special relativity on these grounds.
The absence of a privileged class of equivalent frames (inertial frames) suggests that, in de Sitter relativity it would
be desirable, to characterize significant physical quantities in an intrinsic way, namely in a manner
independent of the choice of any particular coordinate patch.
In this talk we would like to show how this can be accomplished for any set of independent
conserved quantities along the geodesic motion of a free de Sitter
particle. These quantities allow for a natural discussion
of classical pointlike scattering and decay processes.

\end{abstract}

{\small \it
\noindent $^*$This is a short review mainly based on two papers (\cite{1,2}) written in collaboration with A. Kamenshchik, U. Moschella and V. Gorini,
which has been presented to the XVIII SIGRAV Conference, Cosenza (Italy), September 22-25, 2009.}

\section{Introduction}
There are various reasons to consider a de Sitter manifold as a physical background. One is that it is the simplest
solution of the Einstein equations in vacuum, in presence of a positive cosmological constant. The actual cosmological
models must take account of the ``dark energy'' responsible for the accelerating expansion of the universe, revealed
by recent observations. The simplest model for dark energy is just the introduction of a cosmological constant. There are
many possible origins for the cosmological constant, but one should not exclude that, at least in part,
it could be a fundamental constant on the same footing as the velocity of light. Again, this is what happen
for a (anti) de Sitter spacetime, whose isometry group, de de Sitter group, is a two parameter group with the parameters fixed, in the
physical interpretation, by the values $c$ of the speed of light and $\Lambda$ of the cosmological constant.
When adopting such a point of view, one usually prefer to decouple the nature of the universal constants from a
specific physical phenomenon. For example, in Einstein's special relativity the connection between the value of $c$ and
the speed of light is accidental or better, it is a consequence of the nature of Maxwell equations, which describe a
relativistic theory. But Einstein's special relativity does not need of Maxwell's theory to be formulated. \\
After the pioneering paper by Ignatowski \cite{Ignat}, many authors attempted, with success, to deduce special relativity
in a pure geometrical way without appealing to the constancy of the value of the speed of light, se for example
\cite{Frank,Lalan,LevyLeblond1,Gorini,Sen,Giulini,Silagadze}. The better expression of this independence is
in the work f Minkowski, who recognized the Poincar\`e group as the group of a unique entity unifying space and
time into spacetime. What Minkowski missed to do \cite{Dyson} was to extend his point of view so to deduce the (anti) de Sitter
as a possibility for spacetime. However,
it is remarkable to note that the problem to determine all kinematical groups compatible with the basic observational spacetime symmetries
of space isotropy, spacetime homogeneity and boost invariance was solved by Bacry and L\'{e}vy-Leblond in 1968 \cite{Bacry}.
They showed that indeed such group is just the (anti) de Sitter group.
Here we will recall the main steps, slightly generalized in \cite{1}, which lead to a so beautiful general result. \\
A further motivation for studying de Sitter relativistic physics is that it provides the simplest curved background over which
one can define quantum field theories in a rigorous way \cite{M,BEM,BGM}. In order to have insights into the quantum physics
on a de Sitter space it can be very useful to learn new point of view on the description of classical physics on such
backgrounds. We started such a program in \cite{2}, and our main results will be reviewed here. We will consider the problem
of characterizing the constants of motion of massive free particles in a de Sitter background $dS_4$, in a way suitable for
describing elementary processes as collisions and decays. Starting from the embedding of $dS_4$ in a flat Minkowski ambient spacetime,
this will be obtained in terms of certain two forms, living in the ambient space, which evaluated on timelike geodesics
give rise to additive constants of motion. At any point of the worldline, one can introduced locally inertial coordinates such that
the constant are essentially related to the value of the energy-momentum vector at the given point in such coordinates. An advantage
of our formulation is that the quantities are defined in a way independent on the choice of any local coordinate system. In particular, it
allows an intrinsic definition of the relative energy of a free particle w.r.t. a reference timelike geodesic.

\section{The Bacry L\'evy-Leblond construction}

In a remarkable paper \cite{Bacry}, Bacry and L\'evy-Leblond obtained a large class of kinematical groups assuming that space-time is a
homogeneous four-dimensional continuum, spatially isotropic, and invariant under boosts. They also assumed that
space reflection and time reversal are automorphisms of the algebra. However, this last hypothesis can be weakened \cite{1}.
Let us call $J_i, P_i, H$ and $K_i$ ($i=1,2,3$) respectively the
generators of space rotations, space translations, time translation and boosts which we assume to
form a basis for the Lie algebra of the group. Spatial
isotropy implies that {\bf {J}, {P}} and {\bf K} are vectors and that $H$ is a scalar so that the following Lie brackets
must be imposed
\begin{eqnarray}
&& \left[J_i , H \right] =0 \ ,\quad  \left[J_i , P_j \right] =\epsilon_{ijl} P_l \ ,
\quad \left[J_i , J_j \right] =\epsilon_{ijl} J_l \ , \quad \left[J_i , K_j \right] =\epsilon_{ijl} K_l \label{cinque} \ ,
\end{eqnarray}
whereas all remaining brackets will be
\begin{eqnarray}
&& \left[H , P_i \right] =\omega_i H +\gamma_{il} P_l +\varepsilon_{il} J_l +\alpha_{il} K_l \ , \cr
&& \left[H , K_i \right] =\chi_i H +\lambda_{il} P_l +\zeta_{il} J_l +\eta_{il} K_l \ , \cr
&& \left[P_i , P_j \right] =\iota_{ij} H + \varphi_{ijl} P_l +\beta_{ijl} J_l +\psi_{ijl} K_l \ , \label{10} \\
&& \left[K_i , K_j \right] =\xi_{ij} H + \nu_{ijl} P_l +\mu_{ijl} J_l +\upsilon_{ijl} K_l \ ,  \cr
&& \left[P_i , K_j \right] =\rho_{ij} H + \pi_{ijl} P_l+\sigma_{ijl} J_l +\tau_{ijl} K_l \ . \nonumber
\end{eqnarray}
These brackets must obviously satisfy the Jacobi identities. Their imposition constraints the constants to satisfy certain relations,
which we do not report here, see \cite{1}. Beyond this, we simply implement the action of spatial reflections on the generators
${P_i, J_j}$ and ${K_k}$:
\begin{eqnarray}
(P_i,J_j,K_k) \longrightarrow (- P_i, J_j, -K_k)\ .
\end{eqnarray}
This further simplify the relations between the various constants. The most general solution of the resulting constraints, compatible
with noncompactness of boosts, gives
\begin{eqnarray}
&& \left[J_i , H \right] =0 \ , \quad\ \left[J_i , P_j \right] =\epsilon_{ijl} P_l \ , \quad\ \left[J_i , J_j \right] =\epsilon_{ijl} J_l \ ,
\quad\ \left[J_i , K_j \right] =\epsilon_{ijl} K_l \ , \cr
&& \left[H , P_i \right] =\gamma P_i +\alpha K_i \ , \quad\ \left[H , K_i \right] =\lambda P_i -\gamma K_i \ , \quad\
\left[P_i , P_j \right] =\alpha \rho \epsilon_{ijl} J_l \ , \label{Icaso}  \\
&& \left[K_i , K_j \right] =-\lambda \rho  \epsilon_{ijl} J_l  \ ,  \quad\ \left[P_i , K_j \right] =\rho\delta_{ij} H -\gamma \rho
\epsilon_{ijl} J_l \ . \nonumber
\end{eqnarray}
These brackets are invariant under the symmetry
\begin{eqnarray}
S:\left\{ { P_i} \leftrightarrow {K_i}\ , \ \alpha
\leftrightarrow \lambda\ , \ \gamma \leftrightarrow -\gamma\ , \
 \rho \leftrightarrow -\rho\ \right\} \ ,\label{esse}
\end{eqnarray}
which can be used to restrict to the case $\rho>0$.
Moreover, we can reduce ourselves to the case $\gamma=0$ by means of a nonsingular transformation
\begin{eqnarray}
\left( \begin{array}{c} P'_i \\ K'_i \end{array} \right)=
\left( \begin{array}{cc} a & b \\ c & d \end{array} \right)
\left( \begin{array}{c} P_i \\ K_i \end{array} \right),
\qquad\ \left( \begin{array}{cc} a & b \\ c & d \end{array} \right)\in GL(2,\mathbb{R}).
\end{eqnarray}
The final form of the Brackets is then
\begin{eqnarray}
&& \left[J_i , H \right] =0 \ , \quad\ \left[J_i , P_j \right]=\epsilon_{ijl} P_l \ , \quad\ \left[J_i , J_j \right] =\epsilon_{ijl} J_l \ , \cr
&& \left[J_i ,K_j \right] =\epsilon_{ijl} K_l \ , \quad\ \left[H ,P_i \right] =\alpha K_i \ , \quad\ \left[H , K_i \right] =\lambda P_i \ , \label{IIcaso}\\
&& \left[P_i , P_j \right] =\alpha\rho \epsilon_{ijl} J_l \ , \quad\ \left[K_i , K_j \right] =-\lambda\rho \epsilon_{ijl} J_l \ , \quad\
\left[P_i , K_j \right] =\rho\delta_{ij} H \ . \nonumber
\end{eqnarray}
The case $\lambda<0$ must be excluded because it corresponds to have compact boosts. The only possibilities are then classified
by the sign of $\alpha$. For $\alpha >0$ we get the group $SO(1,4)$ which is the symmetry group for the de Sitter space.
For $\alpha <0$ we get the group $SO(2,3)$ which is the symmetry group for the anti de Sitter space.\\
Following observational considerations, we concentrate on the de Sitter case.
To realize the spacetime as a quotient manifold we should look at the orbits of the $SO(1,4)$ subgroup.
However, there is a very simple way to obtain the desired result \cite{1}. Fix any spacelike vector $X\in M^{1,4}$. Then,
$X\cdot X =-R^2$ for some $R>0$ and the matrix
\begin{equation}
M^A_{\ B} (X)= -\delta^A_{\ B} -\frac 2{R^2} X^A X_B
\end{equation}
lies in $SO(1,4)$. It is also held fixed by the $SO(1,3)$ subgroup which fixes the point $X$:
\begin{equation}
AX=X \Longrightarrow A M A^{-1}=M.
\end{equation}
This means that the map $J:X\hookrightarrow M(X)$ gives an embedding of the manifold
$$
W=\{X\in M^{1,4} : X\cdot X =-R^2 \}.
$$
into the group $SO(1,4)$. $W$ inherits a natural metric from the Minkowski metric $\eta$ of $M^{1,4}$, $ds^2_W=\eta|_W$
and $SO(1,4)$ is its isometry group. It is easy
to see that, a part from an armless multiplicative constant, such metric is simply the pullback $J^*(K)$ of the Killing metric
on $SO(1,4)$.\\
Note that the more physical connection among (\ref{IIcaso}) and the symmetry group of the manifold $(W,J^*(K)) $ is
given by the identifications (see \cite{1})
\begin{equation}
\alpha=\frac {c^2}{R^2}, \qquad \lambda=1, \qquad \rho=\frac 1{c^2},
\end{equation}
where $c$ is the speed of light and $R$ is the de Sitter radius, related to the cosmological constant by the relation
$\Lambda=3/R^2$.\\
Before Bacry and L\'evy-Leblond (BLL), much work has been done to deduce the kinematical group of an homogeneous and spatially
isotropic spacetime in various geometrical ways, avoiding the postulate of the absoluteness of the speed of light in vacuum,
starting from the paper of Ignatowsky \cite{Ignat}.
All of them bring to the Poincar\'e group of isometries for a Minkowski spacetime. What is then the point where the BLL approach
overcomes the Ignatowsky theorem leading to more general kinematical groups? The point is that in all previous works it is
assumed, often without mention it, the existence of global frames where the assumed symmetry properties are explicitly
realized. This means that one can choose coordinates which reflect the invariant properties under the action of the symmetry
group. Working in coordinates then one quickly arrive to the Poincar\'e group. However, the symmetries of spacetime do not
require to be manifest in any global reference frame.\footnote{For example, a round sphere $S^2$ is obviously an homogeneous and isotropic
but there do not exist coordinates on it from which the metric result to be independent.}
The advantage of the BLL method is that it does not make use of coordinates but it works abstractly with the algebra. This permit
to include symmetric spaces with non vanishing curvature.\\
Indeed, it happens that the implementation of discrete symmetries, as for example the spatial reflection as we done, is quite
restrictive in the sense that it reduces the number of parameters describing the resulting family of kinematical groups in a significative way.
It should be interesting to study the most general family excluding all discrete subgroups actions on the algebra. These could find
a physical interpretation in terms of deformed Lorentz algebras in quantum gravity approaches, see for example \cite{Girelli:2007xn}.

\section{Conserved quantities}
We will now consider the motion of a massive free particle on the de Sitter space
\begin{equation}
{W\equiv \rm dS}_4=\{X\in M^{1,4}\mid \eta_{AB} X^A X^B=-R^2 \},
\end{equation}
$\eta={\rm diag} \{1, -1,\ldots, -1 \}$. In order to give a global description, in place to
introduce local coordinates on $W$ we will exploit its embedding in $M^{1,4}$.
This can be done by introducing a Lagrange multiplier $\alpha$, so that the usual action for
a particle particle of mass $m$ freely moving on $W$ can be written as
\begin{eqnarray}
S=-mc\int \left[(\eta_{AB}V^A V^B)^{\frac{1}{2}} + \alpha (X^2+R^2)\right]\ d\lambda.\label{freeaction}
\end{eqnarray}
Here $\lambda$ is a parameter for the timelike curves $\lambda \to X^A(\lambda)$ so that the respective
velocity is $V^A(\lambda)={dX^A}/{d\lambda}$. The multiplier enforces the constraint $X^2(\lambda) = -R^2$.
Note that the constraint implies the tangentiality condition $X\cdot V=0$, which
has to be imposed also on the initial conditions.
We see that the isometries of the manifold are symmetries of the action. These are generated by ten
Killing vector fields
\begin{eqnarray}
{\mathcal X}_{AB}=\left.\left( X_A \frac {\partial}{\partial X^B} -X_B \frac {\partial}{\partial X^A} \right)\right|_{dS}.
\end{eqnarray}
Note that the vector fields
\begin{eqnarray*}
H:=\frac cR {\mathcal X}_{04}, \qquad P_i:=\frac 1R {\mathcal X}_{i4}, \qquad K_i=\frac 1c {\mathcal X}_{0i},
\qquad J_i:=\frac 12 \epsilon_{ijk} {\mathcal X}^{jk},
\end{eqnarray*}
satisfy exactly the same Lie brackets as (\ref{IIcaso}). Introducing the five-momenta $\Pi^A:= mdX^A/d\tau$, where $\tau$ is
the proper time for the particle, the conserved quantities associated to the Killing vector fields by means of the
Noether theorem are
\begin{eqnarray}
K_{AB}&=& \frac 1R (X_A \Pi_B -X_B \Pi_A) . \label{duesei}
\end{eqnarray}
These quantities are well defined on any curve, end are independent on $\tau$ when evaluated on geodesic curves.
The quantities $K_{AB}$ have a nice geometrical interpretation: they define a two form dual to the plane generated in
$M^{1,4}$ by the vectors $X^A$, $\Pi^B$. Their conservation along geodesic curves then means that such geodesics are
intersections between the dual plane and the manifold $W$.\\
Not all the obtained conserved quantities are independent. They are completely determined by the initial position
$X^A$, constrained by the condition $X^A X^B \eta_{AB}=-R^2$, and the momenta $\Pi^A$ constrained by the mass shell
condition $\Pi^A \Pi^B \eta_{AB}=m^2 c^2$. Fixing the initial point and takeing into account the condition
$\Pi^A X_B=0$ we see that there are only three fully nontrivial constants of motion, just like for the Minkowskian case.\\
The causal structure of the de Sitter manifold is characterized by the future asymptotic cone
\begin{eqnarray}
C^+ = \{X\in W: \ X^A X^B \eta_{AB}=0, \ X^0>0\}.
\end{eqnarray}
The plane identifying a given timelike geodesic intersects the cone along two generatrices. These are generated by
two future directed null vectors\footnote{w.r.t. the fivedimensional geometric structure} $\xi$ and $\eta$.
Thus, the geodesic can be characterized in terms of these two vectors: parameterizing with the proper time one gets
\begin{equation}
X (\tau) = R \frac { \xi\, e^{\frac {c\tau}R}-\eta \,e^{-\frac {c\tau}R}}{\sqrt {2\xi\cdot\eta }}\ .\label{parametrizzazione}
\end{equation}
Note that there is not a natural choice for the null vectors. They can be changed by positive scale factors, so that we should
equally choose the vectors $\xi'=\mu^2 \xi$, $\eta'=\nu^2 \eta$. This is equivalent simply to a shift of the proper time.
Indeed,
\begin{eqnarray*}
X' (\tau) := R \frac { \xi'\, e^{\frac {c\tau}R}-\eta' \,e^{-\frac {c\tau}R}}{\sqrt {2\xi'\cdot\eta' }}
=R \frac {(\mu/\nu) \xi\, e^{\frac {c\tau}R}-(\nu/\mu) \eta \,e^{-\frac {c\tau}R}}{\sqrt {2\xi\cdot\eta }}=X(\tau+\tau_0),
\end{eqnarray*}
with $\tau_0=(R/c) \log (\mu/\nu)$. Fixing an initial time leaves a free whole rescaling $\mu=\nu$. A complete fixing can then be
obtained by fixing a normalization for the product $\xi \cdot \eta$. We will choose a suitable normalization later.
Using this parametrization, the constants of motion take the form
\begin{eqnarray}
K_{AB}=mc \frac {\xi_A \eta_B-\eta_A\xi_B}{\xi\cdot\eta} \ .\label{massconserved}
\end{eqnarray}
These can be interpreted as defining a two-form $K_{(\xi,\eta)}$ over $M^{1,4}$ such that
\begin{eqnarray}
K_{(\xi,\eta)}: M^{1,4}\times M^{1,4}  \longrightarrow \mathbb{R}, \ (\xi',\eta')\longmapsto mc
\frac {(\xi\cdot \xi')(\eta\cdot \eta')-(\xi\cdot \eta')(\eta\cdot \xi')}{\xi\cdot\eta}.
\end{eqnarray}
We will see that such interpretation of the constants of motion is suitable for describing decay and collision
processes in de Sitter spacetime, without referring to the introduction of local coordinates. In particular it
allows for an intrinsic definition of relative energy.

\section{Relative energy.}\label{sec:energy}
In constructing the kinematical group we discussed the fact that in general the symmetry
group does not manifest itself in any local coordinates system. If one enforces such
requirement than the kinematical group reduces itself to the Poincar\'e group and
the manifold becomes the Minkowski's spacetime. In this case then it exists a preferred
class of frames where the kinematical symmetries are manifest, which is the class of
inertial frames. This is indeed Einstein's special relativity.
As the inertial frames in Einstein's special relativity are global frames, it is then possible
to define end measure the energy of a particle relatively to the frame. But for more general
symmetric spaces there are not preferred global frames and usually the energy of a particle, which is frame
dependent, can be defined only locally.

Our construction for the constants of motion in de Sitter can be used to provide a definition
of the energy of a poinlike particle relative to a reference (massive) free particle representing
a localized observer which should be intended to be at rest. The advantage is that we do not need
to introduce a local frame.

To start with, let us recall that time translations are algebraically generated by the Killing vector field $H:=\frac cR {\mathcal X}_{04}$.
If a pointlike free particle of mass $m$ is moving along the timelike geodesic specified by the null vectors
$\xi, \eta$, this suggests to define its energy as
\begin{eqnarray}
E=cK_{04}=cK_{(\xi,\eta)} (e_0,e_4),
\end{eqnarray}
where $\{e_I\}_{I=0}^4$ is a canonical (inertial) frame for $M^{1,4}$.
Let us introduce the light-like five-vectors $u=e_0+e_4$ and $v=e_0-e_4$. According to our discussion in the previous section, the
select the geodesic passing through the point $O\equiv(0,0,0,0,R)$ with zero velocity, at $\tau=0$. Moreover, we have
\begin{eqnarray}
E=-\frac {cK_{(\xi,\eta)}(u,v)}{u\cdot v}=:E_{(u,v)}(\xi,\eta).\label{energymass}
\end{eqnarray}
Thus, we can interpret this formula by saying that it represent the energy of the massive free particle $m, (\xi, \eta)$ relative to
the reference geodesic $(u,v)$. This interpretation is improved by the observation that we can act on the system with a symmetry
transformation $T\in SO(1,4)$. In the active point of view, this will transform the couples of vectors $(\xi, \eta)$ and $(u, v)$
in new couples $(\xi', \eta')$ and $(u', v')$ representing two different geodesics, so that
\begin{eqnarray}
E=E'=E_{(u',v')}(\xi',\eta').
\end{eqnarray}
This evidences the relativity of $E$.
Note that $E_{(u,v)}(\xi,\eta)=E_{(\xi,\eta)}(u,v)$ which can be interpreted as the symmetry between the active and passive point of
view. Note also that the proper energy is $E_{(\xi,\eta)}(\xi,\eta)=mc^2 $, as it should be. \\
We can give further alternative expressions for the relative energy.
The first one is by noting that we simply have
\begin{eqnarray}
E_{(u,v)}(\xi,\eta)=K_{(u,v)}(\Pi(\tau), X(\tau)),
\end{eqnarray}
where the momenta $\Pi=dX/d\tau$ are computed along the geodesic $X(\tau)\equiv (\xi,\eta)$.
Another interesting hint comes from the curvature Riemann tensor which for the de Sitter manifold result to be
\begin{eqnarray}
{\pmb R}\equiv R_{\mu\nu\rho\sigma}=\frac 1{R^2} (g_{\mu\rho}g_{\nu\sigma}-g_{\mu\sigma}g_{\nu\rho})=
\frac 1{R^2} (\eta_{AC}\eta_{BD}-\eta_{AD}\eta_{BC})\partial_\mu X^A \partial_\nu X^B \partial_\rho X^C \partial_\sigma X^D,
\end{eqnarray}
where $X^A(x^\mu)$ represents the embedding $i: dS_4 \hookrightarrow M^{1,4}$ in some local coordinates. If we introduce the
five dimensional tensor
\begin{eqnarray}
{\pmb \Omega}\equiv \Omega_{ABCD}:=\frac 1{R^2} (\eta_{AC}\eta_{BD}-\eta_{AD}\eta_{BC}),
\end{eqnarray}
then we can write ${\pmb R}=i^* {\pmb \Omega}$. It is convenient to se ${\pmb \Omega}$ as section
of $T^* M^{1,4}\wedge T^* M^{1,4} \otimes \Omega^2 (M^{1,4})$, with
\begin{eqnarray}
{\bf \Omega}: (X^A,Y^B)\longmapsto {\bf \Omega}(X,Y)_{AB}=\frac 1{R^2} (X_A Y_B-Y_A X_B).
\end{eqnarray}
Indeed, if
\begin{equation}
Y(\tau) = R\frac { u\, e^{\frac {c\tau}R}-v \,e^{-\frac {c\tau}R}}{\sqrt{2u\cdot v}}
\end{equation}
is the parametrization of the reference geodesic, we get\footnote{here we use the natural identification $T_p M^{1,4}\simeq M^{1,4}$}
\begin{eqnarray}
{\pmb\Omega}(Y, \dot Y)=K_{(u,v)},
\end{eqnarray}
where the dot is derivative w.r.t. $\tau$.
Then we can also write
\begin{eqnarray}
E_{(u,v)} (\xi,\eta)={\pmb\Omega}(Y, \dot Y) (\Pi(\tau), X(\tau))\equiv {\pmb\Omega}(Y(\tau), \dot Y(\tau),\Pi(\tau), X(\tau)).\label{curvenergy}
\end{eqnarray}
Our definition of energy can be also related to the local energy defined in any given coordinate patch.
Let us introduce some local coordinates $x^\mu$, so that the embedding $i: dS_4 \hookrightarrow M^{1,4}$
is defined locally by the functions $X^I(x^\mu)$. As usual, we mean $x^0=ct$, where $c$ is the speed of light.
For simplicity, we can assume that the origin $(0,0,0,0)$ of these coordinates identify the point $O\equiv (0,0,0,0,R)$
of the hyperboloid. Then, we define the energy $E$ of a free massive particle relative to the given
frame as the energy of the particle w.r.t. the reference geodesic $x^\mu(t)$ passing through the origin with zero velocity:
$x^i(0)= 0$, $\frac {d x^i}{dt} (0)= 0$, $i=1,2,3$. \\
Usually, the energy of a pointlike particle in a local frame is identified with the zeroth component of a four vector. This
is not however the case in general for or conserved relative energy, as we will see in a few examples. The main point is that
there is not any analogous intrinsic definition for the linear momentum.
A definition of momentum for a de Sitter particle necessarily requires the selection of some coordinate system.
One can try to work out some plausible definitions, which however remain related to the choice of local coordinate.
We will not consider any further discussion on it here, see \cite{1}, \cite{2}.

\section{Examples}
We will now consider some examples where the relative energy is compared with the usual notion of local energy in in a local frame.

\subsection{Flat coordinates.}

The flat coordinate system $\{t,x^i\}$ is defined by the equations
\begin{eqnarray}
X(t,x^i)=\left\{\begin{array}{lll}
X^0&=&  R\sinh \frac {ct}R +\frac {\vec x^2}{2R} e^{\frac {ct}R} ,\\
X^i &=&  e^{\frac {ct}R} x^i ,  \\
X^4 &=&  R\cosh \frac {ct}R -\frac {\vec x^2}{2R} e^{\frac {ct}R}.
\end{array}\right. \label{flat-coor}
\end{eqnarray}
In these coordinates the metric for the de Sitter geometry is that of a flat exponentially expanding Friedmann universe
\begin{eqnarray}
ds^2= c^2 dt^2 -e^{ {2ct}/R} \delta_{ij} dx^i dx^j = c^2 dt^2 -a^2(t) \delta_{ij} dx^i dx^j .
\end{eqnarray}
Note that this are yet compatible with the condition $X(0,0,0,0)=(0,0,0,0,R)$. The condition $dx^i/dt=0$ for $t=0$ thus
selects $(u,v)$ with $u=(1,0,0,0,1)$ and $v=(1,0,0,0,-1)$.
The explicit expression for the conserved energy is then
\begin{eqnarray}
&& E=
{mc^2}\frac {dt}{d\tau} -\frac c R x^i p_i= \frac {mc^2}{\sqrt {1-a^2(t) \frac {v^2}{c^2}}}
-\frac c R x^i p_i \label{flatenergy}
\end{eqnarray}
where we have set
\begin{eqnarray}
v^i=\frac {dx^i}{dt}\, ,
\qquad p_i =-m{e^{2 {ct}/R}}\frac {dx^i}{d\tau}=-\frac {m{a^2(t)}v^i}{\sqrt {1-a^2(t) \frac {v^2}{c^2}}} \ .
\label{momenta}
\end{eqnarray}
We introduced the quantities
\begin{eqnarray}
p^i=-\frac{1}{a^2(t)}p_i =
\frac{mv^i}{\sqrt{1-a^2(t)\frac{v^2}{c^2}}} \label{momenta1}
\end{eqnarray}
because as we will see in a moment they can be interpreted as the de Sitter version of the linear momentum.
It is easy to see that in the limit $R\longrightarrow \infty$, (\ref{flatenergy}) and (\ref{momenta1}) go over into the usual
Minkowskian expressions of the energy and momentum.

To compare the obtained result for the energy with the usual definition in local coordinate, let us
now solve the problem of finding the conserved energy associated to a free massive particle in a more
conventional way, by working directly with coordinates.
The action of the point particle in the de Sitter background expressed in the flat coordinates is
\begin{eqnarray}
S=-mc \int \sqrt {1- e^{\frac{2ct}R} \frac {v^i v^j}{c^2} \delta_{ij}}\ dt. \label{freeaction1}
\end{eqnarray}
Note that this action is not invariant under a $t$ coordinate shift, because the local expression for the background
metric is not. However, we can easily find a symmetry associated to time translation takeing account of the spacetime
expansion.
Indeed, in flat coordinates the spatial distances dilate in the course of time by the exponential factor
$e^{\frac{ct}{R}}$, so that compensating it, the expression of an infinitesimal symmetry under time evolution is
\begin{eqnarray}
&&t \longrightarrow t + \epsilon, \nonumber \\
&&x^i \longrightarrow x^i - \frac{c}{R}x^i\epsilon.
\label{time-trans}
\end{eqnarray}
We will identify such transformation with a time translation (in place of a simple $t$ shift).
The action (\ref{freeaction1}) is invariant under such transformation so that we can apply
Noether's theorem to compute the corresponding constant quantity, which results to be precisely
(\ref{flatenergy}).\\
We can note also that the action (\ref{freeaction1}) is manifestly invariant under space translations. Thus we can
also determine the conserved quantities associated to the spatial translations
\begin{equation}
x^i \longrightarrow x^i+\epsilon^i.
\end{equation}
These are exactly the covariant components of the momentum $p_i$ introduced above, so corroborating our
interpretations. In particular, this analysis shows what we anticipated: the conserved energy is not the zeroth component
of a four vector. This is because it is not associate to a simple coordinate shift, but to the transformation (\ref{time-trans}).
If we introduce the tetra momentum
$$
\pi^\mu:= m\frac {dx^\mu}{d\tau},
$$
then we see that
\begin{eqnarray}
E=c\pi^0-\frac c R x^i p_i.
\end{eqnarray}

\subsection{Spherical coordinates.}

Spherical coordinates $\{t, \omega^\alpha \}$, $\alpha=1,\ldots,4$, are such that
\begin{eqnarray}
\left\{\begin{array}{lll}
X^0 &=& R\sinh \frac {ct}R \ ,\\
X^\alpha&=&  R\cosh \frac {ct}R \, \omega^\alpha \ ,
\end{array}\right.
\end{eqnarray}
with $\omega^\alpha \omega^\beta \delta_{\alpha\beta} =1$, which means that $\omega^\alpha$ is a vector on the sphere $S^3$ of
unit radius. Concretely
\begin{eqnarray}
\left\{\begin{array}{lll}
\omega^1&=&\sin \chi^1 \sin \chi^2 \cos \chi^3\\
\omega^2&=&\sin \chi^1 \sin \chi^2 \sin \chi^3 \ ,\\
\omega^3&=&\sin \chi^1 \cos \chi^2 \ ,\\
\omega^4 &=&\cos \chi^1\
\end{array}\right.
\end{eqnarray}
This coordinate system covers a dense subset of the whole de Sitter manifold. The
geometry is that of a closed Friedmann universe undergoing an epoch of exponential contraction which is followed by an epoch of
exponential expansion:
\begin{equation}
ds^2=c^2 dt^2 -R^2 \cosh^2 \frac {ct}R \left\{ \left(d{\chi^1}\right)^2
+\sin^2 \chi^1 \left[\left(d{\chi^2}\right)^2 +\sin^2 \chi^2 \left(d{\chi^3}\right)^2\right]\right\}.
\end{equation}
Again, the null vectors characterizing the associated reference geodesic are $u=(1,0,0,0,1)$ and $v=(1,0,0,0,-1)$.
By using (\ref{curvenergy}) to compute the conserved energy we obtain
\begin{eqnarray}
&& E=mc^2 \frac {\omega^4-\frac R{2c} v^4 \sinh \frac {2ct}R}{\sqrt {1-\frac {R^2}{c^2} \cosh^2 \frac {ct}R v^\alpha v^\beta \delta_{\alpha\beta}}},
\label{ballenergy}
\end{eqnarray}
where $\alpha,\beta=1,2,3,4$ and $v^\alpha=d\omega^\alpha/dt$.\\
Again, this expression for the energy can be  recovered as the
conserved quantity associated to a time translation compensated by a suitable
rescaling of the $\omega$'s in order to leave invariant the
action
\begin{eqnarray}
S=-mc^2 \int \sqrt{1 -\frac {R^2}{c^2} \cosh^2 \frac {ct}R \Omega_{ij} w^i w^j}\ dt \ ,
\end{eqnarray}
where $d\Omega^2 = \Omega_{ij} d\chi^i d\chi^j$ and  $w^i=\frac {d\chi^i}{dt}$.

\subsection{Static coordinates.}
This is an interesting example because it coincide with the coordinate system originally introduced by de Sitter in
his 1917 paper \cite{desitter}. They are defined by
\begin{eqnarray}
\left\{\begin{array}{lll}
X^0 &=& R \sqrt{1-\frac {r^2}{R^2} } \sinh \frac {ct}R,\\
X^i&=& r^i \ , \quad i=1,2,3, \\
X^4&=& R \sqrt{1-\frac {r^2}{R^2} } \cosh \frac {ct}R,
\end{array}\right.\end{eqnarray}
where $r^2=\sum_{i=1}^3 r^i r^i$. With these coordinates the metric exhibits a bifurcate Killing horizon at $r=R$
\begin{eqnarray}
ds^2 =\left( 1-\frac {r^2}{R^2}\right) c^2 dt^2 -\frac {dr^2}{\left( 1-\frac {r^2}{R^2}\right)}
-r^2 (d\theta^2 +\sin^2 \theta d\phi^2) \ .
\end{eqnarray}
For this reason they are also called the black hole coordinates.
As for the previous examples, the associated reference geodesic is characterized by the lightlike vectors
$u=(1,0,0,0,1)$ and $v=(1,0,0,0,-1)$.
The corresponding conserved energy results to be
\begin{eqnarray}
E=mc^2 \left( 1-\frac {r^2}{R^2}\right) \frac 1{\sqrt{1-\frac
{r^2}{R^2}-\frac {(\vec{r}\cdot\dot{\vec{r}})^2}{(R^2-r^2)c^2}
-\frac {\dot {\vec r} \cdot \dot {\vec r}}{c^2} }}.\label{blackenergy}
\end{eqnarray}
Let us look at the action for a massive free particle in these coordinates
\begin{eqnarray}
S=-mc^2 \int \sqrt{1-\frac {r^2}{R^2}-\frac
{(\vec{r}\cdot\dot{\vec{r}})^2}{(R^2-r^2)c^2} -\frac {\dot {\vec
r} \cdot \dot {\vec r}}{c^2}   }\ dt\ .
\end{eqnarray}
Because of the metric is static, the action is independent on the time coordinate $t$, so that this time
the symmetry is just represented by $t$-shifts. The associated conserved energy
coincides with (\ref{blackenergy}).

\section{Collisions and decays.}\label{sec:collisions}
Up to now, we looked specifically at the energy. Here, we will show that the wall set of conserved quantities expressed
in terms of the quadratic form are suitable for describing elementary processes like collisions and decays.\\
Let us consider the case of two particles which collide giving rise to the production of an arbitrary number of outgoing particles
\begin{eqnarray}
a_1 + a_2\longrightarrow b_1+ b_2+\ldots +b_N \ .
\end{eqnarray}
Each ingoing particle $b_i$, with mass $m_{a_i}$, is described by a geodesic curve ending at the collision point $X_0$, which for
clarity we will assume to be the origin point $(0,0,0,0,R)$. As we are describing a purely classical phenomenon, where only pointwise interactions
are considered, $X_0$ is also the starting point of the $N$ geodesics describing the outgoing particles $b_f$ having masses  $\tilde m_{a_f}$.
We can choose the null vectors describing such geodesics in such the way that the collision point will coincide with the common zero of the
proper time of all particles involved in the process, so that $X_0=X_i(0)=X_f(0)$, $i=1,2$, $f=1,2,\ldots,N$.
We will call $(\chi_i, \zeta_i)$ and $(\xi_f,\eta_f)$ the pairs of normalized null vectors identifying the timelike geodesics for the ingoing
and outgoing particles respectively.  \\
To determine the collision law, we can first work with generic local coordinate system $x^{\mu}$,
$\mu=0,1,2,3$. Let us introduce the corresponding four-momenta $\pi^{\mu}=m\frac {dx^{\mu}}{d\tau}$
$\pi^{\mu}=kc\frac {dx^{\mu}}{d\sigma}$) for any given massive particle of timelike worldline $x^{\mu}(\tau)$ on $dS$.
The collision is then described by the conservation of the total covariant energy-momentum four-vector
\begin{eqnarray}
\pi_{a_1}^\mu + \pi^{\mu}_{a_2} =\sum_{f=1}^M \pi^{\mu}_{b_f}. \label{totalmomentum}
\end{eqnarray}
We want to reexpress this condition in terms of the conserved two forms. This is easily done exploiting the fact that they are conserved
and that at the collision point the following relation holds true for each particle:
\begin{eqnarray}
\left.K_{AB}\right|_{X=X_0}=\frac 1R \left(X_{A}
\left.\frac {\partial X_{B}}{\partial x^{\mu}}-X_{B} \frac {\partial
X_{A}}{\partial x^{\mu}}\right)\right|_{x=\bar{x}} \pi^{\mu},\label{Kp}
\end{eqnarray}
where $X_A=X_A(x^\mu(\tau))$. This is nothing but a consequence of the embedding in the five dimensional Minkowski spacetime.
By summing over all ingoing and outgoing particles and using (\ref{totalmomentum}) we get
\begin{eqnarray}
K_{a_1}+K_{a_2} = \sum_{f=1}^N K_{b_f}, \label{scattering}
\end{eqnarray}
which is the desired reformulation for the collision law.
In a similar way, for the decay $a\longrightarrow b_1+b_2+\ldots+b_N$ of a massive particle $a$ we have
\begin{eqnarray}
K=\sum_{f=1}^N K_{b_f}\ . \label{decay}
\end{eqnarray}
Now, solving the collision problem amounts to finding the outgoing  geodesics $(\xi_f,\eta_f)$ given the ingoing ones
$(\chi_i,\zeta_i)$. Apparently, this seems a quite complicated task because the constants $K_{AB}$ are nonlinear
functions of the defining vectors:
\begin{eqnarray}
K_{a_i}=m_{a_i}c \frac {\chi_i \wedge \zeta_i}{\chi_i \cdot \zeta_i}, \qquad\ K_{b_f}=m_{b_f}c \frac {\xi_f \wedge \eta_f}{\xi_f \cdot \eta_f}.
\end{eqnarray}
However, recall that our choice of the lightlike vectors performed so that all proper times of the particle coincide to $\tau=0$ at the
collision instant, fixe each pair of vectors apart from a global rescaling factor (for each pair). We are then free to fix such rescaling factors
so that all products $\chi_i \cdot \zeta_i/m^2_{a_i}$ and $\xi_f \cdot \eta_f/m^2_{b_f}$ are all normalized to the same value, say $2/M^2$.
The condition $X_0=X_i(0)=X_f(0)$ gives
\begin{eqnarray}
&& \zeta_{a_i}=\chi_{a_i} -2\frac {m_{a_i}}{M}\frac {X_0}R, \\
&& \xi_{b_f}=\xi_{b_f} -2\frac {m_{b_f}}{M}\frac {X_0}R.
\end{eqnarray}
Thus the quantities conserved along the single geodesics take the form
\begin{eqnarray}
K_{a_i} =\frac {M c}R X_0\wedge \chi_{a_i}\ ,\ i=1,2\
;\qquad\ K_{b_f}=\frac {M c}R X_0\wedge \xi_{b_f}\ ,\ f=1,2,\ldots,N\ .
\label{33}
\end{eqnarray}
Putting these in the scattering and decay equations (\ref{scattering}) and (\ref{decay})
respectively, we get
\begin{eqnarray}
&& (\chi_{a_1}+\chi_{a_2} -\sum_{f=1}^N \xi_{b_f} )\wedge X_0=0, \label{38}\\
&& (\chi -\sum_{f=1}^N \xi_{b_f} )\wedge X_0=0\ . \label{39}
\end{eqnarray}
The solution of this equations is now an easy problem as equations (\ref{scattering}) and (\ref{38}) are evidently equivalent to
equation (\ref{totalmomentum}). However, they have the advantage of being expressed in an intrinsic form.
To look better at this equivalence, let us make explicit the collision point $X_0=(0,0,0,0,R)$ so that, for example, equation (\ref{38}) becomes
\begin{eqnarray}
\chi_1^\mu+\chi_2^\mu=\sum_{f=1}^N \xi_f^\mu, \ \mu=0,1,2,3. \label{311}
\end{eqnarray}
Now, we know that
\begin{eqnarray}
\zeta^\mu_{i} =\chi^\mu_{i}\ \;\;\mu=0,1,2,3, \mbox{    and    }\
\zeta^4_{i}=\chi^4_{i} -\frac {2m_{a_i}}M.  \label{312}
\end{eqnarray}
Using the fact that $\chi$ and $\zeta$ are null vectors, these relations imply $\chi_{i}^4=-\zeta_{i}^4 =\frac {m_{a_i}}M$ so that we get
\begin{eqnarray}
&& \chi_{i} =\left(\chi^0_i, \vec \chi_i, \frac {m_{a_i}}M \right)\ , \;\;\; \zeta_i=\left(\chi_i^0, \vec \chi_i, -\frac {m_{a_i}}M
\right), \label{313}
\end{eqnarray}
and
\begin{eqnarray}
(\chi_i^0)^2 -(\vec \chi_i )^2 =\frac {m_{a_i}^2}{M^2}. \label{314}
\end{eqnarray}
Using these expression in the parametrization of the geodesics we find that
\begin{eqnarray}
\left. m_{a_i} \frac {dX_{a_i}^\mu}{d\tau}\right|_{\tau=0} =Mc\chi_{a_i}^\mu =:q_i^\mu , \qquad  \left.
m_{a_i}\frac {dX_{a_i}^4}{d\tau}\right|_{\tau=0}=0,\label{315}
\end{eqnarray}
and
\begin{eqnarray} q_i^2=(q_i^0)^2 -(\vec q_i )^2 =m_{a_i}^2 c^2 \ .
\end{eqnarray}
Obviously a similar analysis can be repeated for the outgoing particles.\\
Now, the equivalence is clear. At the collision point, we can choose local coordinates
$x^\mu=X^\mu$, $\mu = 0,1,2,3$. It is then easy to show that $x^\mu$ are locally inertial
coordinates in the sense that in this coordinates the spacetime metric is the flat one at the collision point.
Moreover, the corresponding covariant momenta are exactly the $q_i^\mu$.\\
Thus we conclude that, as it should do, the decay and collision laws in de Sitter even in the intrinsic form
are exactly equivalent to the usual total energy-momentum conservation. This was indeed our
starting point. Curvature does not play any role in classical pointwise interaction, according to the
fact that any Lorentzian manifold is locally inertial. The important advantage, we insist, is that
the expression (\ref{scattering}) and (\ref{decay}) are independent from the coordinates and from the choice
of the collision point, which need to be specified to make explicit the equivalence.

However, it is interesting to note that when quantum effects are considered, the interaction delocalizes and the cosmological constant
will enter in some way in the collision law. Let us consider a very naive reasoning about this working in
flat coordinates, end consider for example the decay of a single particle of mass $m$ into two particles of masses $m_1$
and $m_2$. If we specialize to energy, the decay law take the form
\begin{eqnarray}
c\pi^0-\frac c R x^i \pi_i=c\pi^0_{1}-\frac c R x_1^i \pi_{1i}+c\pi_2^0-\frac c R x_2^i \pi_{2i}.
\end{eqnarray}
At the collision point $x(0)=x_1(0)=x_2(0)$ and the relation reduces exactly to the usual total four-momentum conservation law.
In particular this gives the usual mass law
$$
m\geq m_1+m_2.
$$
But let us now consider the inclusion of quantum effect so that at the collision point the expression
$\frac c R x^i \pi_i-\frac c R x_1^i \pi_{1i}-\frac c R x_2^i \pi_{2i}=0$ will be naively substituted by the Heisenberg
uncertainty principle
$$
\frac c R x^i \pi_i-\frac c R x_1^i \pi_{1i}-\frac c R x_2^i \pi_{2i}\approx \frac cR \hbar.
$$
Then, the conservation law is substituted by the relation
\begin{eqnarray}
\pi^0-\pi^0_1-\pi^0_2\approx \frac \hbar{R},
\end{eqnarray}
so that, for example, the mass condition can be violated.
The decay of a massive scalar particle in de Sitter has been studied rigorously in \cite{M}
where it has been shown that the decay amplitude depends on $\Lambda=\sqrt{\frac 3{R^2}}$ and the presence of
curvature allows in some cases for a non-zero probability for an unstable particle of mass $m$ to decay into particles whose total
mass is larger than $m$.

\section{Detecting particles}
We conclude with another interesting application of our expressions for the conserved quantities. Indeed, note that
the conservation of the total energy-momentum four-vector holds just in the collision point, as on a non flat spacetime
the sum of momenta makes sense only when all of them lie in the same tangent space. Differently from the flat case, the momentum
of each particle is covariantly constant but not constant, because there is a nontrivial connection.
On the other side, in a collision process, or in a decay, the outgoing products are detected far away from the collision point and
the measured momenta are then not directly the ones entering in the conservation law. One must then be able to reconnect the measured
quantities to the ones at the collision point, for example, by solving the geodesics equations and transporting back the momenta.\\
The point is now that the conserved quantities $K_{AB}$, as we seen, are directly connected to the explicit solution of the geodesic
equations so that it is plausible to use them to solve the problem. Let us suppose to work in a given local coordinates frame $x^\mu$
and to measure the momentum $\pi^\mu$ of a free moving particle. Then, we know that this is related to the conserved quantities
by
\begin{eqnarray}
K_{AB}= \left. K_{AB}\right|_{x=x_1} = \left. \frac 1R \left(X_{A} \frac {\partial X_{B}}{\partial x^{\mu}}-X_{B}
\frac {\partial X_{A}}{\partial x^{\mu}}\right)\right|_{x=x_1} \pi^{\mu} (\tau_1)\ ,
\label{Kp1}
\end{eqnarray}
where $x_1=x(\tau_1)$ is the detection point. As $K_{AB}$ are constant and the same expression must be true at the collision
point $x_0=x(\tau_0)$, it is clear that this formula provides a direct connection between $\pi^\mu(\tau_0)$ and $\pi^\mu(\tau_1)$.
However, we do not need to realize such connection if we are interested in the scattering process only. Instead, we can
interpret (\ref{Kp1}) as a measure of $K_{AB}$. If we do that for all the produced particles, than we can use (\ref{scattering})
and (\ref{decay}) in the same way as one uses the total momentum conservation to analyze data in flat spacetime.
Indeed, it is important to note that for a free particle of mass $m$, the mass shell condition generalizes to
$$
2K^2:=K_{AB}K^{AB}=-2m^2 c^2,
$$
which is the other ingredient one needs, for example, to trace Dalitz plots.


\section{Conclusions.}\label{sec:conclusions}
In this paper we reviewed some recent results presented at the ``XVIII Congresso SIGRAV, General Relativity and Gravitational Physics,''
and based on the papers \cite{1} and \cite{2}.

Recent cosmological observations show that the universe is subject to an accelerated expansion:
supernovae observations, ripples in the distribution of galaxies, the apparent size of cold and hot spots in the cosmic microwave background,
the integrated Sachs-Wolfe effect, the study of the growth of galaxy clusters and by the changes in galaxy population of the universe
during the different cosmic epochs, all agree in leading to the conclusion that the universe behaves as it would be permeated
by some sort of gravitationally repulsive energy density. Independently from what exactly is the nature of such energy, the consequent
universe evolution is perfectly compatible with the one obtained by adding a cosmological constant effect to the Einstein equations.\\
We have then assumed the point of view which brings the cosmological constant on the same footing of the limiting speed $c$, t.i. to
be a true classical universal constant (eventually dressed by quantum effects). In a sense, this point of view was yet considered by
Dirac who indeed considered the de Sitter spacetime as the more general kinematic possibility, containing the Minkowski spacetime
as a limit \cite{dirac}. However, the general deduction of this fact from a pure kinematical point of view was first given
in the true remarkable paper of Bacry and L\'evy-Leblond \cite{Bacry}, which unfortunately has been ignored by cosmologists
for a long time. We have reviewed here the beautiful reconstruction of \cite{Bacry}, with some slightest improvement considered
in \cite{1}.\\
This point of view naturally suggests that de Sitter spacetime should replace Minkowski spacetime even in absence of matter. Thus,
the de Sitter symmetry group should replace the Poincar\'e group and special relativity needs to be reformulated on this grounds, giving rise
to a de Sitter special relativity \cite{Al,Guo}. In this context, we have started our program to provide a systematic analysis
of special relativity by considering the dynamics of pointwise interacting particles. The first important difference w.r.t.
Einstein's special relativity, and common to all curved backgrounds, is that in the de Sitter case there not exists any privileged
global coordinate frame as the inertial frames. It has been proposed in \cite{Guo} that the nearest to be inertial systems are provided
by Beltrami's like coordinates, which, however, are not obviously global coordinates. In \cite{2} we have tackled the problem
to define constants of motion for massive and massless free particles in an intrinsic way, independent from the choice of any local
coordinate system. This has been obtained by exploiting the isometric embedding of the maximally symmetric de Sitter manifold into the
five dimensional flat Minkowski spacetime $M^{1,4}$. The de Sitter group is thus just the Lorentz group $SO(1,4)$ of $M^{1,4}$, which
is indeed the subgroup of the isometry group, which leaves the de Sitter hyperboloid invariant. This embedding contains all information
about the manifold. The hyperboloid is asymptotically tangent to a null cone which accounts for the causal structure of the de Sitter
space. From the maximal symmetry it follows that the geodesics are simply given by the intersection between the fourdimensional
spacetime and hyperplanes in $M^{1,4}$ passing through the vertex of the causal cone. Timelike geodesics are in correspondence
with hyperplanes intersecting the cone in two distinct straight lines, and which can represented by (equivalence classes of)
pairs of five-dimensional future directed lightlike vectors $(\xi,\eta)$. These characterize both the geodesics and the
``constant'' two forms $K_{(\xi,\eta)}=k \xi\wedge \eta/(\xi\cdot \eta)$, $k$ being an opportune normalization constant.\\
Beyond giving rise to a intrinsic definition of the constant of motion, this construction permits nice applications to
collisions and decays. This is because in the common event point (the collision point or the decay point) they behave exactly as the
covariant momenta, then obeying the collision law (\ref{scattering}) or the decay law (\ref{decay}). However, differently from
the covariant momenta, in our case the single $K$-quantities associated to each particle are separately conserved, so that the
relation remains true at any time. Thus we have a global characterization of the process, in place of a punctual one. This is
important relatively to the fact that the detection of particles is is not performed simultaneously at the collision point.\\
Moreover, we have seen how the two form $K$ allows to give an intrinsic definition for the energy of a free particle with respect to
a reference timelike geodesic, which can be interpreted as a sharply localized conventionally at rest observer. If $(\xi,\eta)$
are the null vectors selecting the moving free particle and $(u,v)$ the ones parameterizing the observer geodesic, then the relative
energy is $E=K_{(\xi,\eta)}(u,v)$. Here we have also provided a nice relation between the energy (and the constant two forms) to
the curvature tensor of the de Sitter manifold. Next we restricted the definition of the energy to local reference frames thus
comparing it with the usual notion of the energy in local charts.\\
Here we have referred explicitly to massive particles, but all construction we have made are easily extendable to massless particles \cite{2}.
Also, all quantities reduces to usual ones in the flat limit $\Lambda\to 0$ \cite{1}. We remark here that the Bacry and L\'evy-Leblond
construction has not yet carried to the most general situation, as a good simplification as introduced by imposing certain (reasonable)
discrete symmetries. It could be that the most general situation could be useful to introduce certain deformed spacetime symmetries
in the ambit of simple quantum gravity approaches. It is plausible to expect that our methods, based essentially on the highly
symmetric non degenerate\footnote{in any case all the degenerate possibilities can be obtained from the general case by contractions} geometry,
should apply in such cases as well.     \\
The next step should be to extend these intrinsic methods to a quantum mechanical description of relativistic particles as well as to classical
fields on a de Sitter background.

\vskip 1.5 cm
\subsection*{Acknowledgments}
The author is grateful to U. Moschella, M. Francaviglia and L. Lusanna for invitation to the
``XVIII Congresso SIGRAV, General Relativity and Gravitational Physics, Cosenza, 22-25 Settembre 2008''.
He thanks S. Liberati and L. Lusanna for interesting comments during the talk.




\end{document}